# Prediction of Distant Metastasis in Head and Neck Cancer Patients Using Tumor and Peritumoral Multi-Modal Deep Learning

**Nuo Tong[1#], Changhao Liu[2#], Zizhao Tang[1], Feifan Sun[3], Yingping Li[1], Shuiping Gou[1*], Mei Shi[2*]**

[1]School of Artificial Intelligence, Xidian University, Xi'an, Shaanxi, 710071, China

[2]Department of Radiotherapy, Xijing Hospital, Air Force Medical University of PLA, Xi'an, Shaanxi, 710032, China

[3]Department of Oncology, The General Hospital of Western Theater Command, Chengdu, China

Nuo Tong and Changhao Liu contributed equally.

Corresponding Authors:

Mei Shi, Shuiping Gou

mshi82@fmmu.edu.cn, shpgou@mail.xidian.edu.cn

# 1. [Abstract]

**Background** Although the combined treatment of surgery, radiotherapy, chemotherapy, and emerging target therapy has significantly improved the outcomes of patients with head and neck cancer, distant metastasis remains the leading cause of treatment failure. Once distant spread occurs, the prognosis is extremely poor, and reliable tools for predicting of metastatic risk are still lacking. In this study, we proposes a deep learning–based multimodal fusion framework that integrates CT imaging, radiomics, and clinical characteristics to predict metastasis risk in patients with head and neck squamous cell carcinoma (HNSCC).

**Methods** A total of 1497 patients were retrospectively analyzed. Tumor and organ masks were generated from pretreatment CT scans. A 3D Swin Transformer extracted deep imaging features, while 1562 radiomics features were computed and reduced to 36 key variables through correlation filtering and random forest selection. Clinical data, including age, sex, smoking, and alcohol status, were encoded and fused with imaging-derived features. The multimodal representation was fed into a fully connected network for metastasis prediction. Five-fold cross-validation was used to assess performance via AUC, accuracy, sensitivity, and specificity.

**Findings** The multimodal model outperformed all single-modality baselines. The deep learning module alone achieved an AUC of 0.715, whereas multimodal fusion significantly improved prediction performance (AUC = 0.803, ACC = 0.752, SEN = 0.730, SPE = 0.758). Stratified analyses confirmed good generalizability across tumor subtypes. Ablation experiments demonstrated that each modality contributed complementary information, and the 3D Swin Transformer provided more robust feature representation compared with conventional architectures.

**Interpretation** This multimodal deep learning model enables accurate, non-invasive prediction of metastasis risk in HNSCC, offering a comprehensive characterization of tumor biology. The approach shows strong potential as a clinical decision-support tool for individualized treatment planning.

**Funding** National Natural Science Foundation of China (No. 82272735, T2541002, 62372358, 62302355), Key R&D Program of Shaanxi Province (No. 2024GH-ZDXM-35, 2024SF-YBXM-213, 2024SF2-GJHX-34), Sichuan Provincial Natural Science Foundation (No. 2025ZNSFSC0648), Xijing Hospital Medical Staff Technical Enhancement Program (No. 2024XJSY18).

# 2.[Introduction]

Head and neck cancers typically refer to malignant epithelial tumors arising in the nasal cavity and paranasal sinuses, the nasopharynx, the oral cavity, the oropharynx, and the hypopharynx. Most originate from squamous epithelium and are collectively called head and neck squamous cell carcinomas(HNSCC). Among these, nasopharyngeal carcinoma (NPC) is associated with EBV infection and is relatively sensitive to radiotherapy and chemotherapy, hence it is categorized separately. Although surgery followed by adjuvant radiotherapy/chemoradiotherapy for resectable HNSCC, and comprehensive strategies based on concurrent chemoradiotherapy combine with neoadjuvant/adjuvant chemotherapy or targeted therapy for unresectable HNSCC and NPC, have significantly improved locoregional control and survival rates for these patients, the incidence of distant metastasis after treatment still hovers around 20–40% during follow-up[1].Once distant metastasis occurs, the median survival time of patients is only approximately 12–14 months. Therefore, accurately predicting the potential risk of metastasis prior to treatment is of great clinical significance for guiding individualized treatment, avoiding overtreatment, and preventing delays in therapy.

Traditional prediction methods primarily rely on imaging observations (e.g., lymph node size and morphology), histopathological indicators (such as tumor differentiation and depth of invasion), and clinical information (including TNM staging, age, smoking, and alcohol history)[2][3][4][5]. However, these methods often depend heavily on subjective assessments, suffer from poor reproducibility, and lack sensitivity in detecting early or subtle metastatic disease[6]. In recent years, the rapid development of artificial intelligence, particularly deep learning and radiomics, has led to a surge of interest in non-invasive, imaging-based risk prediction approaches. Radiomics enables high-throughput extraction of quantitative features such as texture, shape, and intensity from medical images, revealing latent biological behaviors of tumors beyond human visual perception[7][8][9][10]. Meanwhile, deep learning models, with their powerful capacity for automated feature learning, can extract high-level representations directly from raw images and have demonstrated superior performance in cancer diagnosis and treatment outcome prediction of various cancers[11][12].

Emerging studies have confirmed that integrating multimodal information, including deep learning features, radiomic features, and clinical parameters, to build multimodal predictive models significantly improves stability and generalizability across multiple centers and disease contexts[13]. For instance, in predicting treatment responses for nasopharyngeal and cervical cancers, fusion models have shown higher AUCs and greater robustness compared to unimodal approaches. Furthermore, the manner of feature integration affects the ultimate model performance. Early fusion, which concatenates various features before classification, preserves complementary information from different modalities, while late fusion aggregates outputs from separate models through ensemble learning strategies to enhance discriminative capability[14].

Nevertheless, current multimodal research on metastasis risk prediction in head and neck cancerstill faces several challenges. Firstly, most existing studies focus exclusively on specific subtypes (such as oropharyngeal or laryngeal cancer) and lack evaluation of joint modeling across heterogeneous HNSCC populations. Secondly, many models rely solely on two-dimensional images or handcrafted features, neglecting the three-dimensional spatial structure and its interactions with clinical data. Thirdly, these models often lack interpretability and scalability, hindering their direct application in real-world clinical settings. Thus, there is an urgent need for robust multimodal predictive models capable of integrating diverse imaging and clinical data while delivering high accuracy, generalizability, and interpretability.

In this study, we propose a multimodal fusion framework that integrates 3D Swin Transformer-dervied deep features, radiomic features, and structured clinical variables for metastasis risk prediction in head and neck cancer patients. By combining tumor imaging with patient-specific clinical information, the proposed approach aims to capture complementary biological signals and provide a more comprehensive representation of tumor behavior. The model was trained with five-fold cross-validation and validated on independent public and private datasets to ensure robustness and generalizability. The ultimate goal of this research is to develop a reproducible, interpretable, and clinically applicable risk prediction tool to support precision medicine and intelligent decision-making in the management of HNSCC.

# 3.[Methods]

## 3.1 Ethics

This multicenter retrospective study obtained ethical approval from Ethics Committee and Institutional Review Boards of Xijing Hospital, Air Force Medical University of PLA (NO. KY20223027-1). Due to the retrospective nature of this study, the requirement for informed consent was waived.

## 3.2 Patient population

This study included seven datasets, comprising a total of 1497 patients with Head and Neck Squamous Cell Carcinoma (HNSCC) for model training and performance evaluation. The specific data sources are as follows:

**(1) Nasopharyngeal Carcinoma Cohort#1:** 408 nasopharyngeal carcinoma patients treated at Xijing hospital in China from 2012 to 2022.

**(2) Head-Neck-CT-Atlas Dataset (MD Anderson Cancer Center):** 189 consecutive HNSCC patients who received radical radiotherapy between 2003 and 2013.

**(3) Oropharyngeal Cancer Dataset (MD Anderson Cancer Center):** 410 oropharyngeal cancer patients treated between 2005 and 2012.

**(4) Head-Neck-PET-CT Dataset:** 296 HNSCC patients treated at four different institutions in Quebec between April 2006 and November 2014.

**(5) HEAD-NECK-RADIOMICS-HN Dataset:** 137 HNSCC patients who received radiotherapy, sourced from the Cancer Imaging Archive.

**(6) Nasopharyngeal Carcinoma Cohort#2:** 23 HNSCC patients who received radiotherapy at Hanzhong Central Hospital, Hanzhong, Shaanxi, China.

**(7) Nasopharyngeal Carcinoma Cohort#3:** 34 HNSCC patients who received radiotherapy at the Department of Radiation Oncology and Shandong Provincial Key Laboratory of Radiation Oncology, Shandong Cancer Hospital and Institute, Shandong First Medical University and Shandong Academy of Medical Sciences, Jinan, Shandong, China.

Tumors arising from the hypopharynx, larynx, and epiglottis were collectively categorized as laryngopharyngeal carcinoma for analysis, following common clinical classification principles.

The detailed statistical information for all samples is provided in **Table 1**. All enrolled cases met the following criteria:

**1.**Pathologically confirmed diagnosis of HNSCC;

**2.**Clear indication of distant metastasis or multiple primary cancers;

**3.**Manual delineation of gross tumor volume (GTV) masks by experienced radiation oncologists prior to treatment.

**Table1** Patient Demographic Information

**(A)** Detailed Information of the 5 Training and Validation Sample Sets

| Dataset | (1) | (2) | (3) | (4) | (5) |
|---|---|---|---|---|---|
| Case Count | 408 | 189 | 410 | 296 | 137 |
| Tumor Location | Nasopharynx | Laryngopharynx<br>Nasopharynx<br>Oral cavity<br>Oropharynx | Oropharynx | Laryngopharynx<br>Nasopharynx<br>Oropharynx | Laryngopharynx<br>Oropharynx |
| Metastatic Ratio | 110 (27.0%) | 22 (11.6%) | 33 (8%) | 40 (13.5%) | 8 (5.8%) |
| Age | 290(71.1%) | 160(84.7%) | 356(86.8%) | 225(76.0%) | 111(81.0%) |
| Drink | 48(14-83) | 57(24-91) | 58(28-87) | 63(18-90) | 62(44-83) |
| Smoke | 152(37.3%) | -- | -- | -- | -- |
| Treatment Type | RT 15 (3.7%)<br>CCRT 146(35.8%)<br>IC+CCRT 243(59.6%)<br>IC+RT 4(1.0%) | -- | -- | -- | -- |

**(B)** Number of Samples per Subtype and Metastatic Sample Proportion

| Tumor type | Nasopharynx | Oropharynx | Laryngopharyngeal | Oral cavity | Sinus | Unknown |
|---|---|---|---|---|---|---|
| Case count | 441 | 819 | 135 | 7 | 3 | 34 |
| Metastatic ratio | 119<br>(27.0%) | 70<br>(8.5%) | 19<br>(14.1%) | 1<br>(14.3%) | 1<br>(33.3%) | 3<br>(8.8%) |

**(C)** Detailed Information of the External Test Set Data

| Dataset | (6) | (7) |
|---|---|---|
| Case count | 23 | 34 |
| Tumor type | Nasopharynx | Nasopharynx |
| Metastatic case ratio | 7(30.4%) | 29(85.3%) |
| Gender (Male %) | 17(73.9%) | 25(73.5%) |
| Age(mean[range]) | 53(29-74) | 45(16-77) |
| Drink | 3(13.0%) | 5(14.7%) |
| Smoke | 4(17.4%) | 11(32.4%) |

**Figure 1** shows the flowchart of the included patients in the training, internal validation, and external validation sets. The tumor locations in the datasets include nasopharynx, oropharynx, laryngopharyngeal, oral cavity, and nasal cavity, representing multiple subtypes. The clinical characteristics of patients, such as gender ratio, age distribution, smoking and drinking status, varied across the datasets. The detailed statistical data can be found in **Table 1**. The diverse clinical and anatomical backgrounds across centers provide a solid foundation for ensuring the generalizability of the model.

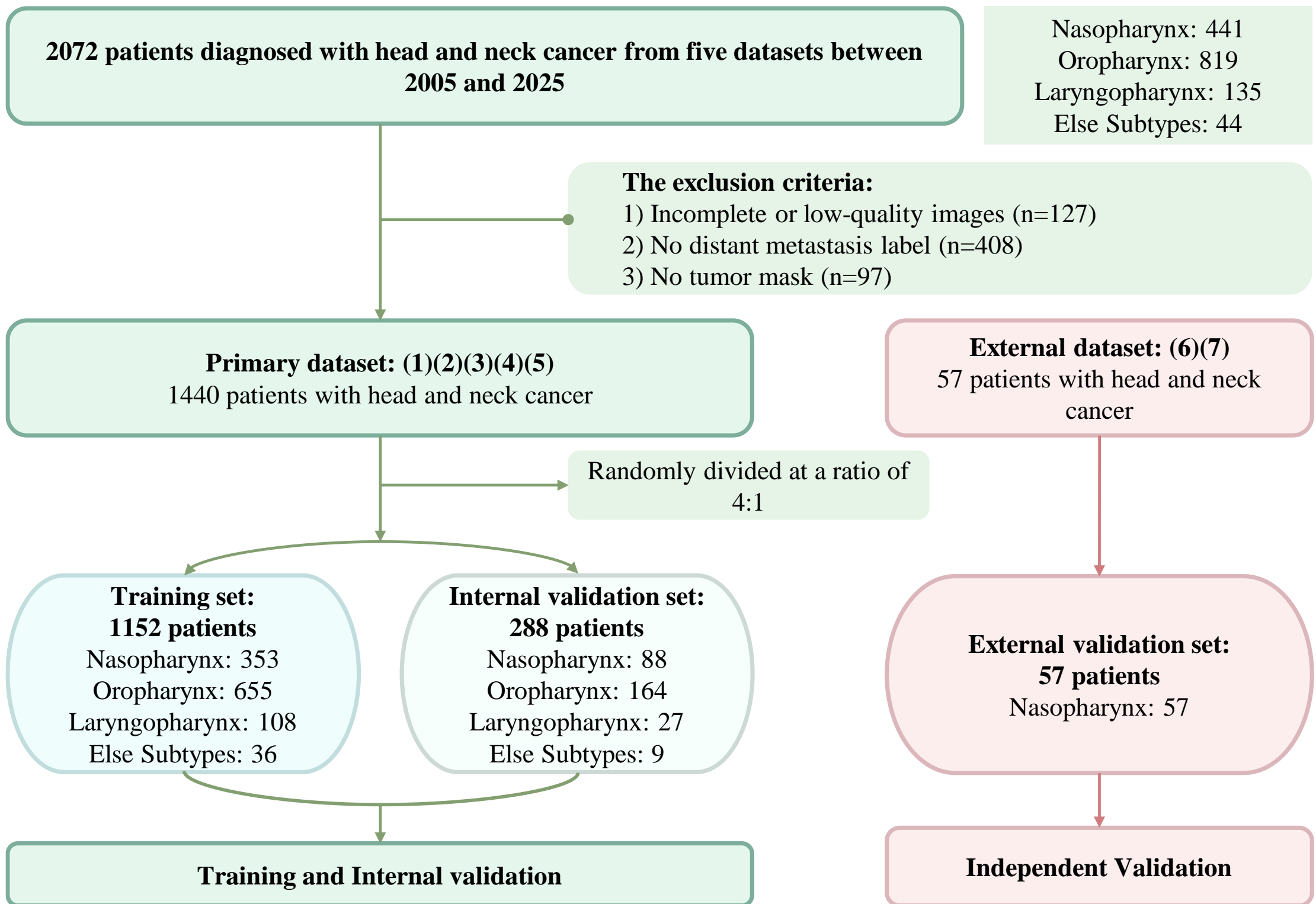

**Figure 1:** Flowchart of the study enrollment process.

## 3.3 CT examination and image preprocessing

The overall study design is illustrated in **Figure 2**. In this study, all patients underwent preoperative scanning using a multi-slice spiral CT system in a supine position, with the scanning range extending from the cranial base to the sternoclavicular joint. Tumors were delineated by experienced physicians to generate tumor masks, and for some patients, masks of nearby organs-at-risk were also available. In subsequent use of the data, for samples without organ masks, the organ regions were set to zero. To standardize the visualization of head and neck anatomy, all CT images were normalized with a window width of 350 and a window level of 50, and the Hounsfield Units (HU) were truncated to a range of [-125, 225] to enhance soft tissue contrast.

Subsequently, all images were resampled spatially: the CT images were interpolated using cubic spline interpolation (B-spline), and the corresponding mask labels were interpolated using nearest neighbor interpolation, with the voxel size unified to 1 mm × 1 mm × 3 mm.

Due to the varying sizes of the original CT images, directly inputting them into the neural network would result in significant memory overhead and hinder model stability during training. Therefore, the spatial distribution boundaries of the tumors and risk organs across all samples were analyzed to determine the bounding box for the entire dataset. Based on this, each patient's CT image and corresponding mask were cropped to ensure uniform input data size, with the region of interest focused on the tumor-related areas.

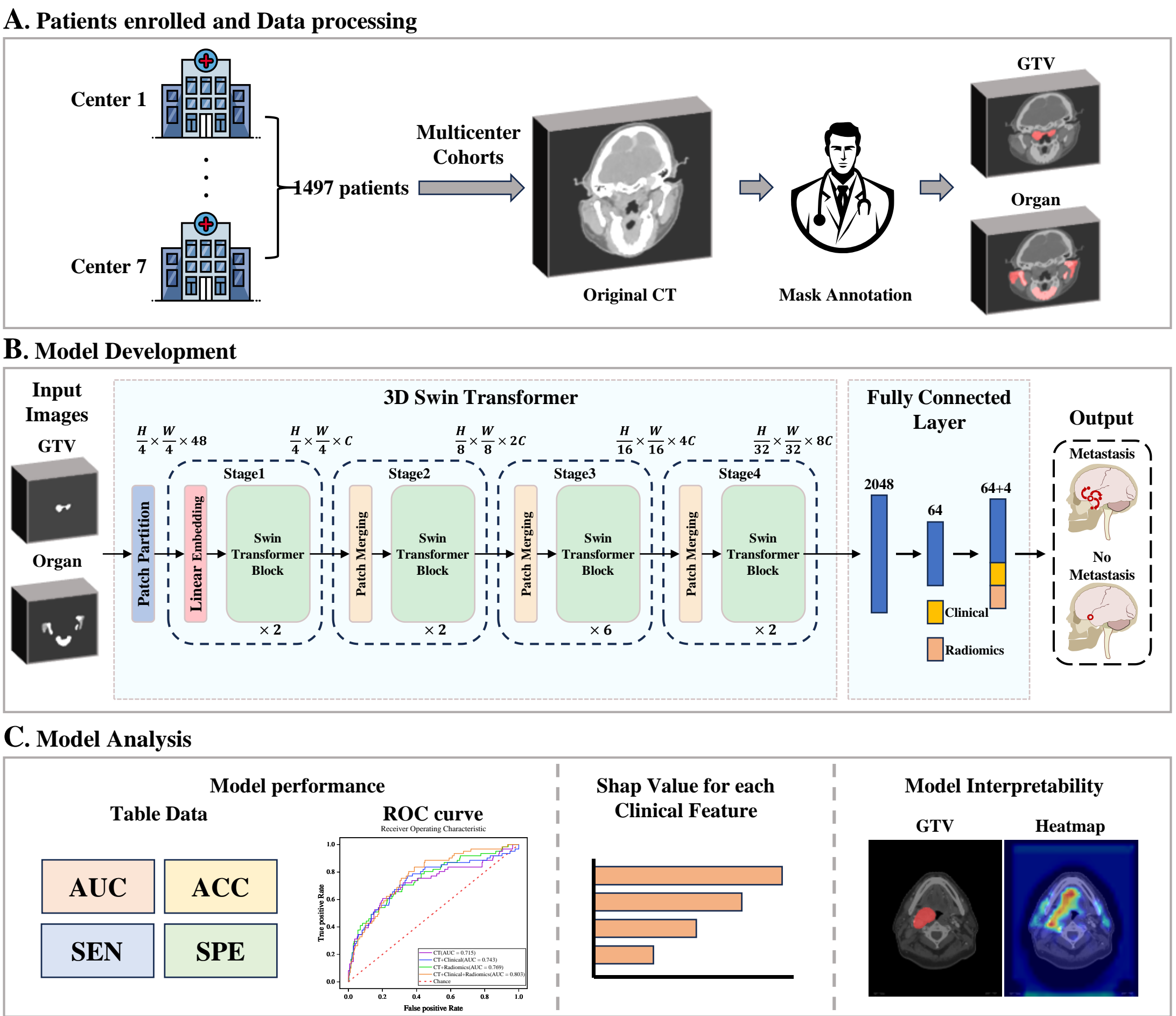

**Figure 2: Overview of the study design.** (A) Data collection and image processing: Patient CT images and clinical records were collected from 7 centers and then assigned to the training, internal validation, and external validation datasets. Manual delineation of gross tumor volume (GTV) masks by experienced radiation oncologists prior to treatment. (B) Model development: Swin Transformer was employed to develop the prediction models. (C) Model assessment: The performance of the model was assessed using ROC, AUC, ACC, SEN, and SPE. The individual decisions made by the model were visualized and interpreted using Grad-CAM.

## 3.4 Radiomics feature extraction

In this study, the Python package PyRadiomics 3.0.1 was used to extract radiomic features from the CT images. The extracted features cover multiple aspects, including shape features, first-order statistics, gray-level co-occurrence matrix (GLCM), gray-level run-length matrix (GLRLM), gray-level dependence matrix (GLDM), and wavelet features, among others. A total of 1,562 features were extracted from each sample[15].

To address potential numerical instability during the model training process, all features were normalized using z-score standardization to eliminate dimensional inconsistencies between different features. Due to the high dimensionality of the extracted features, which could lead to information redundancy, we applied feature reduction methods. First, features with high pairwise correlation were removed through correlation analysis. Then, feature selection was performed using a Random Forest algorithm[16]. With the aid of a random search strategy, a final set of 36 features was retained for each modality to be used in subsequent analysis.

## 3.5 Clinical model construction

Clinical features such as smoking, drinking, age, and gender, as shown in Table 1, were used alongside radiomic features to construct the clinical model. The age feature was standardized using z-score normalization to eliminate dimensional discrepancies. Gender, smoking, and drinking features were encoded using one-hot encoding. Due to missing data for smoking and drinking information in some samples, the corresponding one-hot encoding was divided into three categories: "Yes," "No," and "Missing," ensuring that the model could properly handle incomplete clinical information. The processed clinical features were concatenated with the radiomic features for each sample to form the Clinical model.

## 3.6 3D Deep Learning model development

Convolution-based 3D CNNs are limited in modeling global dependencies and long-range contextual cues. These cues are crucial for metastasis prediction because relevant morphological signals, including subtle peritumoral changes and diffuse infiltration patterns, often extend beyond local receptive fields. To address this issue, we adopted a 3D Swin Transformer[17][18] as the backbone of our framework. Its shifted-window self-attention enables efficient capture of global spatial interactions, and its hierarchical multi-scale design improves the recognition of heterogeneous tumor characteristics that distinguish metastatic from non-metastatic lesions. As illustrated in Figure 1.B, the input volumetric CT images were first divided into non-overlapping 3D patches through a patch partition module, which were then projected into a latent feature space by a linear embedding layer. The embedded tokens were subsequently processed through four hierarchical stages of Swin Transformer blocks. In Stage 1, two Swin Transformer blocks operated on the feature maps of size $\frac{H}{4} \times \frac{W}{4} \times 48$. After patch merging, Stage 2 reduced the resolution to $\frac{H}{8} \times \frac{W}{8}$ and doubled the channel dimension, followed by two Swin Transformer blocks. Stage 3 further downsampled the resolution to $\frac{H}{16} \times \frac{W}{16}$ with quadrupled channels and was processed by six Swin Transformer blocks. In Stage 4, the resolution was reduced to $\frac{H}{32} \times \frac{W}{32}$with eight-fold channels, followed by two Swin Transformer blocks. This hierarchical architecture with shifted-window multi-head self-attention enabled the network to capture both local representations and global spatial dependencies in volumetric medical images. A global average pooling was then applied to generate a 2048-dimensional image feature vector. To adapt the architecture to our binary classification task (metastasis vs. no metastasis), the final prediction head was replaced with a fully connected layer, and the entire network was trained from scratch with randomly initialized weights in an end-to-end manner for 300 epochs.

## 3.7 Construction of the prediction model

To construct the final prediction model, a multimodal fusion strategy was employed in which features from imaging, radiomics, and clinical data were concatenated into a single feature vector. The 2048-dimensional image features extracted by the 3D Swin Transformer were first reduced to 64 dimensions through a fully connected layer. These 64 image features were then combined with four clinical parameters and the radiomic features obtained earlier, forming a joint multimodal representation. The fused feature vector was subsequently passed through another fully connected layer, which compressed it into a one-dimensional output used to generate the final prediction result, where a probability score determined whether the patient was classified as metastasis or no metastasis.

## 3.8 Model performance assessment and interpretability

The performance of each model in predicting distant metastasis was assessed by the area under the curve (AUC) of the receiver operating characteristic (ROC) curve. In addition, accuracy (ACC), sensitivity (SEN), and specificity (SPE) were also calculated to provide a comprehensive evaluation. To further interpret the model's decision-making process, we employed gradient-weighted class activation mapping (Grad-CAM [19]) to generate visual heatmaps. Grad-CAM is a technique that highlights the regions in the input image that contribute most to the model's prediction. In the resulting heatmaps, warmer colors (such as red and yellow) indicate areas where the model focuses more strongly, suggesting that these image regions play a critical role in distinguishing patients with or without distant metastasis. Conversely, cooler colors (such as blue) represent regions with little contribution to the decision. For clinical readers, these heatmaps can be intuitively understood as showing "where the model is looking" when making predictions, thereby providing insights into whether the algorithm's attention aligns with clinically relevant anatomical structures or lesions.

## 3.9 Statistics

All statistical analyses were performed using Python (version 3.12). Continuous variables were summarized as mean ± standard deviation, and categorical variables were expressed as counts and percentages.

Comparisons of the areas under the receiver operating characteristic curves (AUCs) between different models were conducted using the DeLong test. A two-sided P value less than 0.05 was considered statistically significant.

To further interpret the contribution of individual features, SHAP (SHapley Additive exPlanations) values were calculated to quantify the importance of clinical and radiomic variables in the multi-modal fusion model.

# 4.[Results]

### 4.1 Experimental settings

A five-fold cross-validation strategy was used to evaluate model performance. The dataset was evenly divided into five subsets. In each iteration, four subsets were used for training and one for testing, and this procedure was repeated across five rounds to reduce the influence of data partitioning. Each training round consisted of 300 epochs. All experiments were performed on a workstation equipped with an NVIDIA GeForce RTX 3090 (24 GB) GPU to ensure training efficiency and stability.

To comprehensively validate the effectiveness of the proposed multimodal fusion framework, the experimental design included the following components:

**(1) Training and Internal validation**

1. Assessing the effect of different modality combinations (CT images, clinical variables, radiomics features) on metastasis prediction performance.
2. Evaluating the contribution of organ-at-risk information.
3. Quantifying the impact of individual clinical variables, including smoking, alcohol consumption, age, and sex.
4. Examining predictive performance across head and neck cancer subtypes (e.g., nasopharyngeal carcinoma, oropharyngeal carcinoma).
5. Comparisons on various networks.

**(2) External Validation**

External validation was conducted using two independent datasets collected from different regions to assess the model's generalizability.

## 4.2 Assessing the effect of different modality combinations on metastasis prediction performance.

To evaluate the contribution of different types of information in the prediction model, this study compared the predictive performance of three feature combination strategies: CT images + clinical information, CT images + radiomic features, and a fusion of all three, using the five multi-center datasets for model training and validation. The corresponding results are shown in **Figure 3** and **Table 2**.

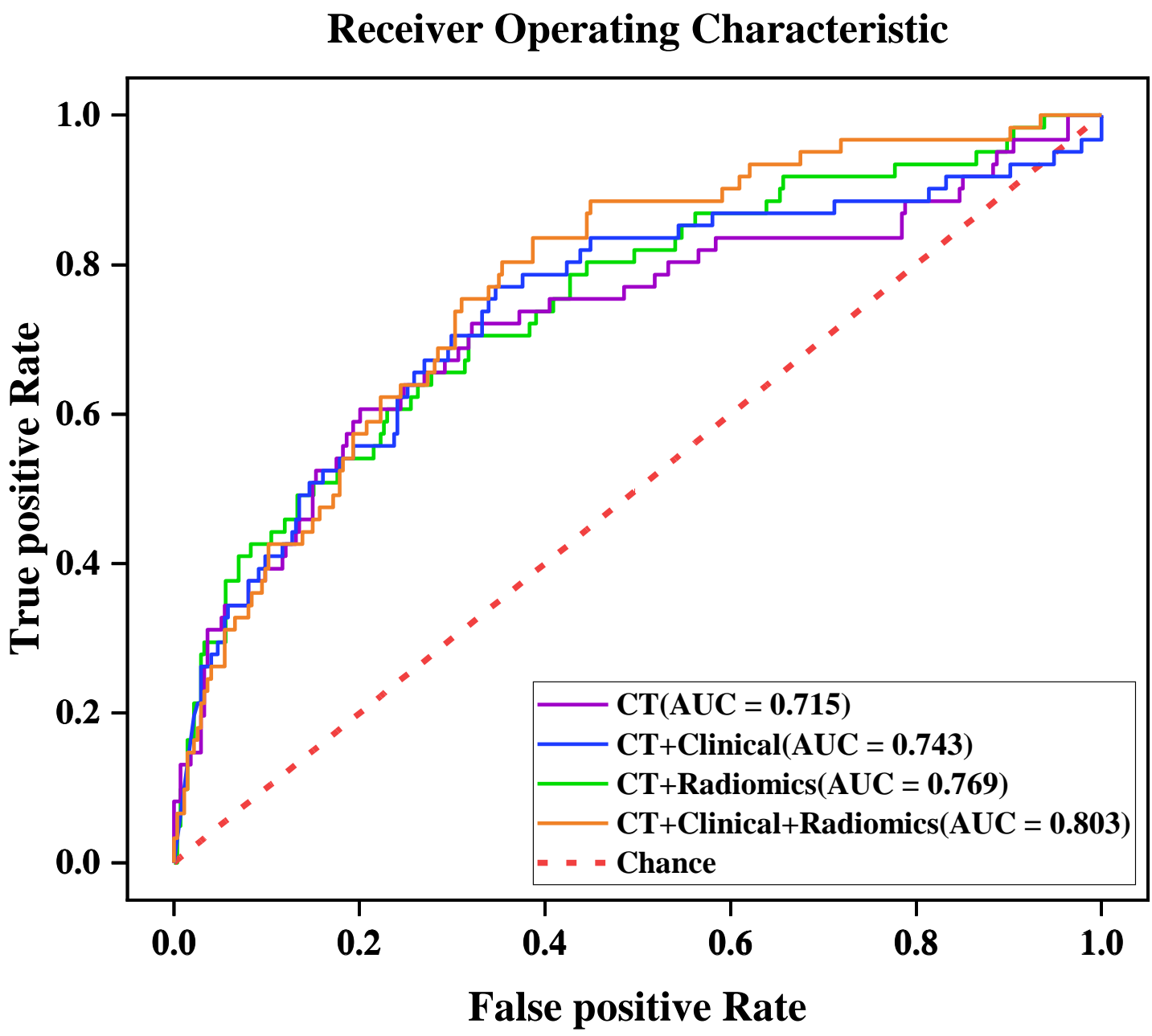

**Figure 3: Predictive performance of different combinations of CT, clinical information, and radiomics.** The receiver operating characteristic (ROC) curves of: (a) CT model, (b) CT + Clinical model, (c) CT + Radiomics model, (d) CT + Clinical + Radiomics model. Abbreviations: ROC, receiver operating characteristic; AUC, area under the curve.

**Table 2**

(A) Performance of different combinations of CT, clinical information ,and radiomics

| Data combination | AUC | ACC | SEN | SPE |
|---|---|---|---|---|
| CT | 0.715 | 0.692 | 0.685 | 0.693 |
| CT+Clinical | 0.743 | 0.731 | 0.718 | 0.733 |
| CT+Radiomics | 0.769 | 0.729 | 0.723 | 0.730 |
| CT+Clinical+Radiomics | **0.803** | **0.752** | **0.730** | **0.758** |

**(B)** Validation of different subtypes using the trained optimal multi-modal model (CT+Clinical+Radiomics)

| **Tumor type** | **AUC** | **ACC** | **SEN** | **SPE** |
|---|---|---|---|---|
| Nasopharynx | 0.724 | 0.705 | 0.667 | 0.719 |
| Oropharynx | **0.822** | **0.774** | **0.786** | **0.773** |
| else | 0.807 | 0.778 | 0.750 | 0.774 |

The results (**Figure 3** and **Table 2(A)**) show that using CT images alone provides preliminary predictive ability (AUC = 0.715), while incorporating structured clinical information (such as age, gender, smoking, and drinking) or radiomic features improves the model's performance to 0.743 and 0.769, respectively, indicating that these two types of information provide effective supplementation to the image model. Notably, when all three types of information are fused, the model's performance improves to its best state (AUC = 0.803), with increased accuracy, sensitivity, and specificity, validating the complementary and synergistic effects of multi-modal features. Statistical comparison using DeLong tests showed that the CT+Clinical+Radiomics model significantly outperformed CT-only, CT+Clinical, and CT+Radiomics models (all $P < 0.05$).

In addition to the results shown in Table 2 (A), we further refine the validation process by breaking down the performance into distinct tumor subtypes, as shown in Table 2 (B). This was done because the previous verification process in Table 2 (A) aggregated all subtypes into a single category, which might have diluted the individual characteristics and predictive capabilities of each subtype. By segmenting the validation into three categories (Nasopharynx, Oropharynx, and the other rare subtypes), we achieve more specific insights into model performance.

The "Else" category represents a consolidation of tumor types that were too sparse in numbers to be treated individually, but nonetheless remain important for the overall model's generalization. The results show that the model performs best for Oropharynx tumors, with an AUC of 0.822, ACC of 0.774, SEN of 0.786, and SPE of 0.773, demonstrating strong predictive performance for this subtype. The Nasopharynx subtype also performs well, with AUC = 0.724, while the "Else" category has a slightly lower performance but still maintains a decent AUC of 0.807.

**Impact of Case Number Differences**

A possible explanation for the varying performance across tumor subtypes is the difference in case numbers. The Oropharynx subtype has 819 cases, nearly twice the number of Nasopharynx cases (441). The larger sample size for Oropharynx tumors provides the model with more data, allowing it to better learn the distinct features associated with this subtype, leading to better generalization and stronger predictive performance. In contrast, the Nasopharynx subtype, with its relatively smaller sample size, may have contributed to the model's less robust generalization ability, as it had fewer data points to capture the underlying tumor characteristics.

By splitting the data in this manner, we can better understand how the model behaves across different tumor types, helping to identify strengths and potential areas for further improvement in subtype-specific predictions. This approach confirms that multi-modal information (CT, clinical, and radiomics) offers consistent benefits, but the exact improvements depend on the specific characteristics of each tumor type. Moreover, the disparity in sample sizes highlights the importance of having a balanced dataset for each subtype to ensure fair training and improve generalization across all tumor types.

### 4.3 Evaluating the contribution of organ-at-risk information

In the preceding experiments, all models were trained using dual-channel inputs, including both the tumor region and the surrounding risk organs, to fully exploit complementary anatomical information. To further validate whether the incorporation of adjacent risk organs indeed contributes to performance improvement, we conducted an ablation experiment on the entire training dataset by using tumor-only inputs.

**Table 3**

Predictive performance of models using tumor-only input versus tumor plus risk organs input.

| Input type | AUC | ACC | SEN | SPE |
|---|---|---|---|---|
| GTV | 0.713 | 0.687 | 0.590 | 0.703 |
| GTV+Organ | **0.803** | **0.752** | **0.730** | **0.758** |

As shown in **Table 3**, the tumor-only model achieved limited predictive performance, with an accuracy (ACC) of 0.687 and an area under the curve (AUC) of 0.713. In contrast, when both the tumor and risk organs were included as input channels, the model's performance substantially improved, reaching an ACC of 0.752 and an AUC of 0.803, and the improvement was statistically significant according to the DeLong test ($p < 0.05$).

These results demonstrate that the inclusion of surrounding risk organs provides additional discriminative information, which plays a critical role in enhancing the model's ability to predict metastasis. From both anatomical and biological perspectives, risk organs adjacent to the tumor are closely associated with potential metastatic pathways, and their integration allows the model to capture more comprehensive features. This design highlights the importance of incorporating peritumoral regions for more reliable metastasis risk prediction.

### 4.4 Quantifying the impact of individual clinical variables, including smoking, alcohol consumption, age, and sex.

To further explore the contribution of different modality information to the model's prediction of tumor metastasis risk, this study conducted multiple modality combination experiments using the **Nasopharyngeal Carcinoma Cohort (Xijing Hospital)**. This dataset contains complete clinical information (including **smoking, drinking, age, gender, treatment type and N stage**), which facilitates the systematic evaluation of the effects of multi-modal fusion strategies. The specific combination strategies, corresponding model performance and contribution of each clinical feature to model prediction are shown in **Table 3 and Figure 4**.

**Table 3**

**(A)**Performance of different combinations of clinical information and CT

| Data combination | AUC | ACC | SEN | SPE |
|---|---|---|---|---|
| CT | 0.693 | 0.657 | 0.636 | 0.664 |
| CT+Smoke | 0.688 | 0.691 | 0.655 | 0.705 |
| CT+Drink | 0.711 | 0.706 | 0.664 | 0.724 |
| CT+Age | 0.699 | 0.701 | 0.600 | 0.738 |
| CT+Gender | 0.693 | 0.694 | 0.600 | 0.728 |
| CT+ N stage | 0.744 | 0.728 | 0.681 | 0.745 |
| CT+ Treatment | 0.634 | 0.610 | 0.627 | 0.604 |
| CT+Smoke+Drink+ Age+Gender+N stage | **0.750** | **0.735** | **0.681** | **0.755** |

**(B)**SHAP Value(contribution of each clinical variable to model prediction performance)

| **Clinical Variable** | **Mean(\|SHAP Value\|)** |
|---|---|
| Smoke | 0.10 |
| Drink | 0.15 |
| Age | 0.30 |
| Gender | 0.38 |
| N stage | **0.57** |

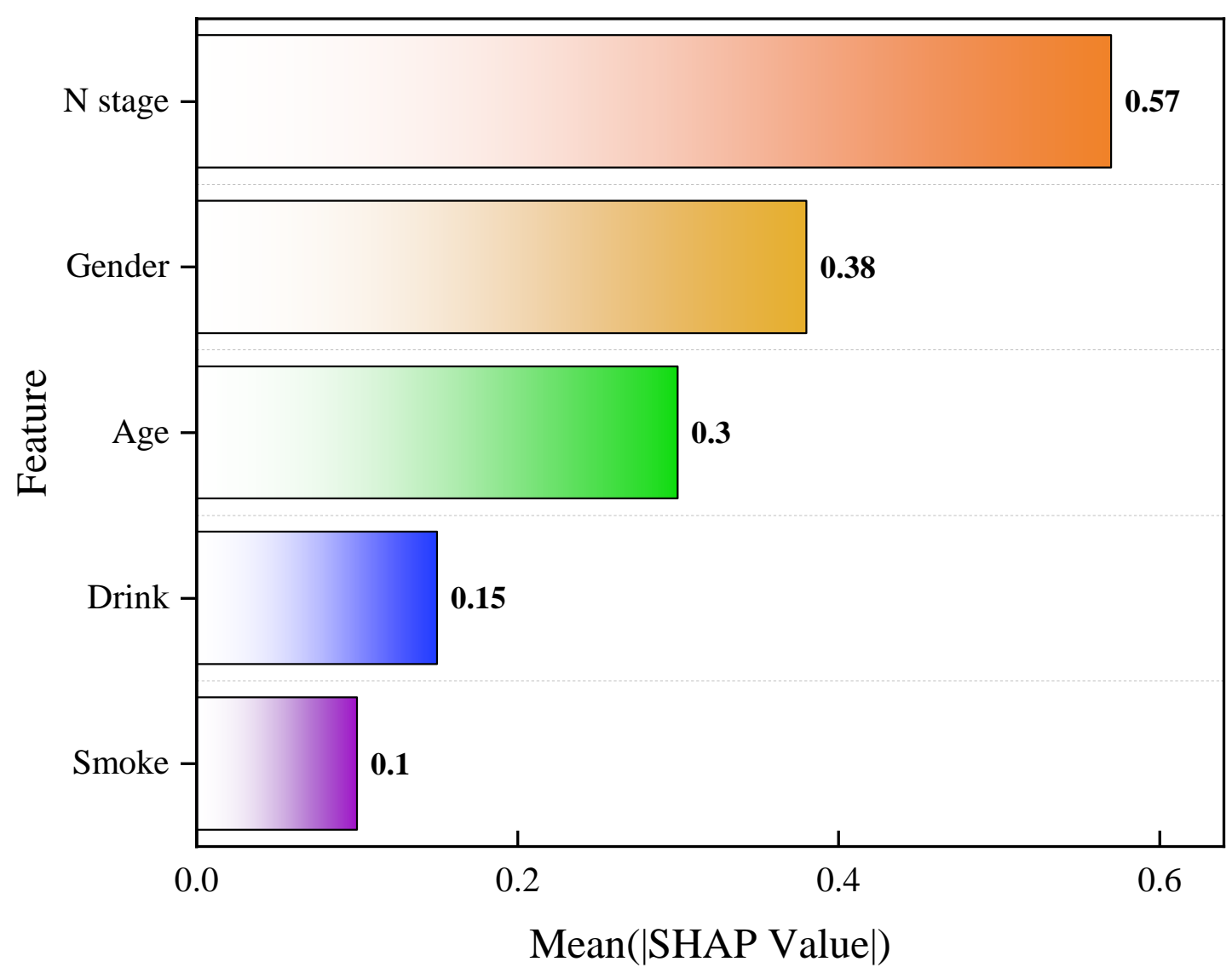

**Figure 4: SHAP Value**
(contribution of each clinical variable to model prediction performance)

As shown in Table 3(A), integrating individual clinical variables with CT image features consistently improved predictive performance compared with using imaging alone. The largest gains were achieved with age and N-stage, with AUC increasing from 0.670 to 0.707 and 0.730, respectively, suggesting a strong association with metastatic risk. Statistical analysis confirmed that these improvements were significant ($P < 0.05$ for all). It is noteworthy that the CT +

treatment type combination performed worse than the model using CT images alone, with AUC dropping to 0.634. This suggests that the "treatment type" variable has a limited contribution to predicting metastatic risk, likely because it reflects clinical decisions rather than tumor biological characteristics.

When the five key clinical variables (smoking, drinking, age, gender, and N-stage) were fused together, the model performance reached its optimal level, with AUC increasing to 0.750 and accuracy (ACC) reaching 0.735. The improvement compared with CT alone was statistically significant ($P < 0.01$). This indicates that the synergy of multi-dimensional clinical information can effectively enhance the model's predictive capability.

The above results suggest that during the model construction process, the reasonable selection and fusion of strongly correlated structured clinical information can provide more discriminative evidence for deep learning models. This is particularly important in medical image tasks with limited sample sizes.

Table 3(B) presents the mean absolute SHAP values[20] for different features in the model. The SHAP values quantify the contribution of each feature to the model's predictions. Features with higher SHAP values have a greater impact on the model's output. According to the table:

N stage has the highest mean SHAP value of 0.57, indicating that it plays the most significant role in predicting metastasis risk.Gender follows with a mean SHAP value of 0.38, showing its considerable but relatively lower influence compared to N stage.Age has a mean SHAP value of 0.30, contributing moderately to the model's predictions.Drink and Smoke features show relatively lower SHAP values of 0.15 and 0.10, respectively, suggesting that their impact on predicting metastasis risk is comparatively less significant.These results highlight the relative importance of clinical variables, particularly N stage and gender, in predicting metastasis risk in Head and Neck Squamous Cell Carcinoma (HNSCC).

### 4.5 Examining predictive performance across head and neck cancer subtypes (e.g., nasopharyngeal carcinoma, oropharyngeal carcinoma).

To evaluate the model's adaptability and generalization across different tumor subtypes, we designed four data configurations using the five internal datasets: nasopharyngeal carcinoma only, oropharyngeal carcinoma only, a combination of both, and the full dataset containing all subtypes. The prediction performance of these configurations is summarized in **Table 4**.

**Table 4**

Performance of different combination strategies for head and neck cancer subtypes

| Data combination | AUC | ACC | SEN | SPE |
|---|---|---|---|---|
| Nasopharynx | 0.678 | 0.646 | 0.630 | 0.652 |
| Oropharynx | 0.670 | 0.660 | 0.600 | 0.666 |
| Nasopharynx+Oropharynx | 0.685 | 0.678 | 0.656 | 0.682 |
| All Data | **0.715** | **0.692** | **0.685** | **0.693** |

The results indicate that when the model was trained using only nasopharyngeal or oropharyngeal carcinoma samples, it exhibited a certain level of predictive capability, but the performance was relatively limited. However, when these two subtypes were combined for training, the model's AUC, accuracy, and specificity showed notable improvement, suggesting that tumors originating from different anatomical sites contain complementary imaging features that, when integrated, enable the model to capture the underlying characteristics of tumor metastasis more comprehensively.

Furthermore, when "other subtypes" (e.g., laryngeal cancer, hypopharyngeal cancer, oral cancer, etc.) were included in the dataset, the model achieved its best performance across all evaluation metrics (AUC = 0.715, ACC = 0.692, SEN = 0.685, SPE = 0.693). DeLong tests confirmed that the multi-subtype model significantly outperformed single-subtype models (all $P < 0.05$). This finding highlights the clinical relevance of incorporating multi-subtype and multi-center data in model development, as it better reflects the heterogeneity of head and neck cancers encountered in real-world practice. By learning from diverse tumor subtypes, the model gains enhanced generalization and robustness, thereby improving its potential utility as a decision-support tool for clinicians in assessing the risk of metastasis.

In conclusion, although single-subtype modeling can yield moderate predictive performance, the integration of multi-subtype data represents a more clinically meaningful strategy. It not only improves model stability and reliability but also aligns more closely with the complexity of actual clinical scenarios, facilitating broader applicability in diagnosis and personalized treatment planning.

## 4.6 Comparisons on Various Networks.

To comprehensively evaluate the effectiveness of different network architectures on multi-modal data, we compared the predictive performance of several widely used 3D convolutional and transformer-based models, including Medical ResNet-34, Medical ResNet-50, Medical ResNet-101, 3D VGG16, 3D DenseNet-121, 3D EfficientNet-B4, 3D ViT, and the proposed 3D Swin Transformer. The results are summarized in **Table 5**.

**Table 5** Comparison of the performance of different models.
(The best results are highlighted in bold, and the second-best results are underlined.)

| model | AUC | ACC | SEN | SPE |
|---|---|---|---|---|
| Medical ResNet-34 | 0.733 | 0.726 | 0.667 | 0.737 |
| **Medical ResNet-50** | 0.788 | **0.767** | 0.713 | **0.779** |
| Medical ResNet-101 | 0.749 | 0.726 | 0.636 | 0.741 |
| 3D VGG16 | 0.703 | 0.667 | 0.613 | 0.687 |
| 3D Dense net-121 | 0.731 | 0.753 | 0.576 | **0.779** |
| 3D EfficientNet-B4 | 0.774 | 0.720 | **0.713** | 0.724 |
| 3D ViT | 0.721 | 0.697 | 0.667 | 0.704 |
| **3D Swin Transformer** | **0.803** | 0.752 | **0.730** | 0.758 |

Overall, transformer-based models exhibited superior performance compared with conventional convolutional networks. Among the ResNet variants, Medical ResNet-50 achieved the best results, with an AUC of 0.788, accuracy of 0.767, sensitivity of 0.713, and specificity of 0.779, outperforming both shallower (ResNet-34) and deeper (ResNet-101) counterparts. This suggests that increasing network depth beyond a certain extent does not necessarily enhance feature representation, and may instead lead to performance degradation due to overfitting.

Compared with the ResNet series, 3D VGG16 showed limited predictive ability (AUC = 0.703), likely attributable to its relatively simple architecture and insufficient capability in capturing long-range dependencies. 3D DenseNet-121 demonstrated improved specificity (0.779) but lower sensitivity (0.576), indicating an imbalance in model prediction performance. Similarly, 3D EfficientNet-B4 achieved a relatively high AUC of 0.774 and sensitivity of 0.713, but its overall accuracy remained lower than that of ResNet-50.

The transformer-based approaches further highlighted the advantages of attention mechanisms in modeling global contextual information. Although the 3D ViT reached only moderate performance (AUC = 0.721, ACC = 0.697), the proposed 3D Swin Transformer achieved the best results across all metrics, with an AUC of 0.803, ACC of 0.752, SEN of 0.730, and SPE of 0.758. These results demonstrate that the hierarchical design and shifted window self-attention mechanism of the Swin Transformer enable more effective feature learning from multi-modal volumetric data, leading to robust and balanced predictive performance.

In summary, the comparative analysis confirms that while conventional CNNs provide a reasonable baseline for metastasis prediction, transformer-based architectures, particularly the 3D Swin Transformer, deliver superior performance and represent a promising direction for future applications in multi-modal medical image analysis.

## 4.7 External Validation Results

To further validate the adaptability and generalization performance of the proposed multi-modal fusion model in real clinical settings, we selected patient data from two independent clinical centers that were not involved in the training process: the "Nasopharyngeal Carcinoma Cohort#2" (23 cases) and the "Nasopharyngeal Carcinoma Cohort#3" (34 cases), for performance evaluation on external test sets. Both datasets are predominantly composed of nasopharyngeal carcinoma cases, with metastasis rates of 30.4% and 85.3%, respectively. Detailed demographic characteristics are shown in Table 6.

**Table 6** External validation results

| Data | AUC | ACC | SEN | SPE |
|---|---|---|---|---|
| Nasopharyngeal Carcinoma Cohort#2 | 0.712 | 0.678 | 0.622 | 0.700 |
| Nasopharyngeal Carcinoma Cohort#3 | 0.698 | 0.651 | 0.645 | 0.667 |

Model's performance on the Nasopharyngeal Carcinoma Cohort#2 was AUC = 0.712, ACC = 0.678, SEN = 0.622, and SPE = 0.700, on the Nasopharyngeal Carcinoma Cohort#3, the performance was AUC = 0.698, ACC = 0.651, SEN = 0.645, and SPE = 0.667. Although the model’s performance slightly declined compared to the multi-center data during the training and validation stages, it still demonstrated strong discriminative ability in both test sets, particularly maintaining high specificity on the Nasopharyngeal Carcinoma Cohort#2. This indicates that the proposed three-modal fusion prediction framework exhibits a certain level of transferability and stability on external datasets, effectively identifying metastatic risks in patients from different sources.

The above results further demonstrate that the deep learning multi-modal model proposed in this study not only performs well in cross-subtype combined modeling and feature fusion strategies but also possesses a certain degree of clinical applicability, making it suitable for metastatic risk prediction tasks in different medical institutions.

## 4.8 Predictive performance stratified by distant metastatic locations

To further investigate the clinical applicability of the proposed model, we analyzed its predictive performance across different distant metastatic sites, including the lung, liver, bone, and brain. The results are summarized in Table 7.

**Table 7 Predictive performance stratified by metastatic regions**

| Metastatic Location | Sample Numbers | Correct Predictions Numbers | Accuracy (%) |
|---|---|---|---|
| Lung | 9 | 4 | 44.4 |
| Liver | 6 | 5 | 83.3 |
| Bone | 12 | 9 | 75.0 |
| Brain | 1 | 1 | 100.0 |

As shown in Table 7, the predictive performance varied across metastatic regions. The model achieved 100.0% accuracy in predicting brain metastasis; however, this result should be interpreted with caution, as only one case was available and therefore may not reflect an objective or generalizable pattern. In contrast, the model demonstrated strong predictive capability for liver (83.3%) and bone metastases (75.0%), while lung metastasis prediction showed relatively lower accuracy (44.4%). This suggests that although the model captured discriminative information for most metastatic sites, challenges remain in lung metastasis prediction, likely due to limited sample size and the high heterogeneity of pulmonary lesions.

Overall, the stratified analysis highlights the potential of the model in capturing metastatic patterns across multiple distant organs, while also pointing out areas for improvement. Expanding the dataset, particularly for lung and brain metastasis cases, may further enhance the model's predictive stability and generalizability in clinical practice.

## 4.7 Model Interpretability

To further elucidate the underlying decision mechanisms of the proposed model, we conducted a comprehensive interpretability analysis combining Grad-CAM visualization and gray-level distribution assessment. Specifically, we compared the spatial attention patterns and intensity histograms between cases with and without distant metastasis.

As shown in Figure5, metastatic cases typically exhibited stronger and more concentrated activations within the gross tumor volume (GTV), suggesting that the model effectively captured discriminative information from biologically aggressive tumor regions. In contrast, non-metastatic cases displayed more diffuse or peripheral activation patterns, indicating a lower degree of feature concentration on the tumor core. Correspondingly, the gray-level histograms revealed that metastatic tumors tended to have a broader intensity distribution, potentially reflecting increased heterogeneity and density within the lesion.

Further comparison among cases with different metastatic sites revealed that metastatic cases generally presented with relatively wider and more intense activation regions within the GTV, accompanied by broader gray-level distributions. In contrast, non-metastatic cases tended to show more localized or peripheral activation patterns, along with narrower gray-level ranges. In other subsets of cases, the Grad-CAM heatmaps often showed concentrated high-intensity activations within subregions of the GTV, bthis pattern was not universal either.

In addition, the gray-level histograms generally indicated that tumors in patients with distant metastasis exhibited broader or multimodal distributions. This suggests increased intratumoral complexity and microenvironmental diversity. Moreover, cases with prominent low-gray (low-density) components were more frequently associated with distant metastasis, which may correspond to necrotic or hypoxic regions that promote angiogenesis and enhance metastatic potential. Collectively, these findings offer a plausible radiobiological interpretation for the model's discriminative focus and highlight its potential to capture clinically meaningful imaging patterns.

Although these preliminary observations highlight meaningful correlations between Grad-CAM attention patterns, gray-level distribution, and metastatic status, further validation using a larger dataset is warranted.

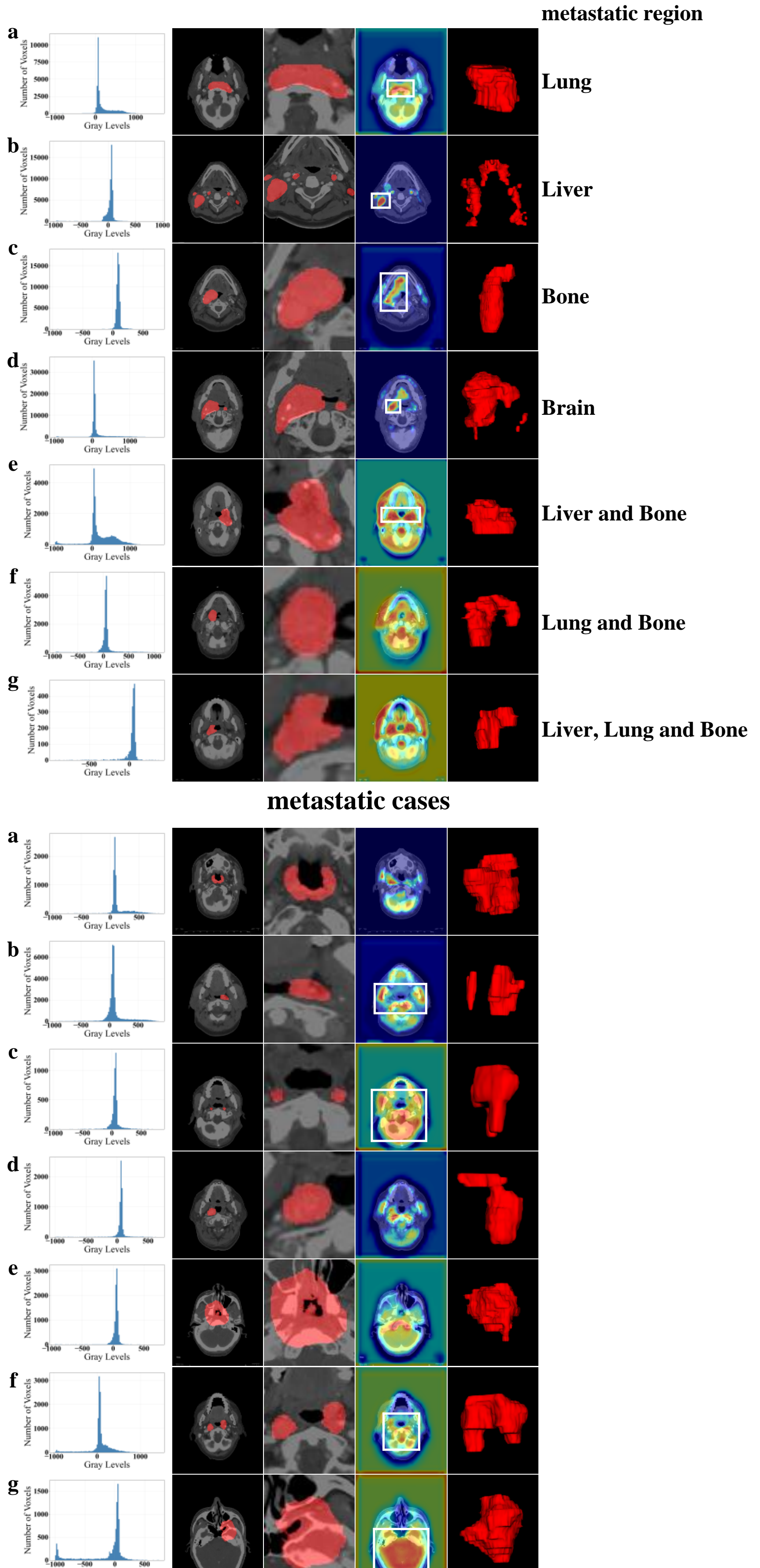

non-metastatic cases

**Figure 5:** Representative heatmaps of correctly predicted HNSCC patients with distant metastasis were generated using Grad-CAM. The visual presentation was organized into five columns, Gray-scale histogram of pixel values in the tumor region, displaying the CT images alongside the enlarged tumor regions, corresponding heatmaps, and 3D tumor volumes. The heatmaps were color-coded to emphasize discriminative regions and were superimposed on the respective ultrasound images. White bounding boxes were added to highlight activated regions, making it easier to distinguish concentrated vs. diffuse/peripheral activation patterns and their potential relevance to metastasis. The prediction model primarily focused on tumor regions. The metastatic regions are shown in the last column.

# 5.[Discussion and Conclusion]

In this feasibility study, we have demonstrated the potential of a deep learning-based multimodal fusion framework for predicting metastasis risk in Head and Neck Squamous Cell Carcinoma (HNSCC) and nasopharyngeal carcinoma (NPC). By integrating CT images, radiomic features, and clinical data, the proposed model significantly outperformed traditional single-modality approaches. The results from the 5-fold cross-validation indicate that the multimodal fusion model not only enhanced predictive performance but also improved generalizability across different subtypes of head and neck cancers, with an AUC of 0.803, accuracy of 0.752, sensitivity of 0.730, and specificity of 0.758.

When compared with models trained on single modalities, the multimodal fusion model demonstrated clear advantages. The incorporation of clinical information, such as age, smoking, and N-stage, along with radiomic features from CT images, provided complementary insights into the metastatic potential of tumors. This is particularly evident in the analysis, where the model trained on combined clinical and imaging data outperformed those trained on either modality alone, with the highest AUC recorded for the fusion of clinical, radiomic, and CT data (AUC = 0.803). The results align with recent studies showing that integrating multimodal information leads to more robust and accurate predictions in cancer prognosis, as it allows the model to capture both the biological behavior of the tumor and the associated clinical risk factors.

The external test set results further validate the model's applicability to diverse patient populations. While performance slightly declined on external datasets, the model maintained strong specificity, especially in the Nasopharyngeal Carcinoma Cohort#2, where it demonstrated an AUC of 0.712 and specificity of 0.700. These results suggest that the model has a reasonable degree of transferability, indicating that it could be applied to clinical practice for predicting metastatic risk in HNSCC patients from different institutions. Nevertheless, future studies with larger sample sizes from a broader range of clinical settings are needed to confirm the model's generalizability and robustness across more diverse patient demographics.

In this study, the internal training and validation cohorts contained multiple head and neck cancer subtypes, whereas both external test cohorts consisted exclusively of nasopharyngeal carcinoma cases. This distribution mismatch may lead to biased feature representation favoring NPC-related radiological patterns, thereby influencing external model performance. Such heterogeneity reflects a real-world challenge, but it also highlights the necessity of subtype-balanced prospective validation across laryngopharyngeal and oral cavity tumors.

The model's high sensitivity is particularly important in the context of medical decision-making, where missing a positive case of metastasis could lead to significant clinical consequences. By providing non-invasive and interpretable predictions, our framework could assist clinicians in making more informed decisions regarding patient treatment plans.

Despite the promising results, this study has limitations. The relatively small sample size of the external test sets, the inclusion of only nasopharyngeal carcinoma, and the absence of a prospective validation cohort are significant limitations that should be addressed in future research. Additionally, the impact of contouring variations on model performance was not assessed, though it is an important factor that could influence the accuracy of predictions. Future work could focus on evaluating the effects of contour variations and incorporating auto-segmentation techniques to mitigate inter-rater variability. As shown in Figure 6, we provide examples of misclassified cases. It can be observed that the model did not accurately focus on the GTV region, which may partly explain the incorrect predictions. Notably, the observed variations in gray-level distributions may reflect certain underlying tumor microenvironment characteristics potentially associated with metastasis risk. However, these findings are preliminary and require further validation on larger cohorts. Future studies should further combine these features with texture-based descriptors for quantitative modeling and analysis, to better capture intratumoral heterogeneity and improve the model's ability to identify clinically relevant regions. This suggests that there is still room for improvement in our approach. Future work could aim to enhance the model's ability to capture more discriminative features within the GTV region, thereby allowing the model to focus more precisely on the clinically relevant areas and improve prediction accuracy.

In conclusion, the proposed deep learning-based multimodal fusion model demonstrates high accuracy and robust performance in predicting metastasis risk in HNSCC, with the potential for clinical application in personalized treatment planning. With further validation and integration of other imaging and clinical parameters, this approach could significantly reduce unnecessary neck dissections and improve patient outcomes.

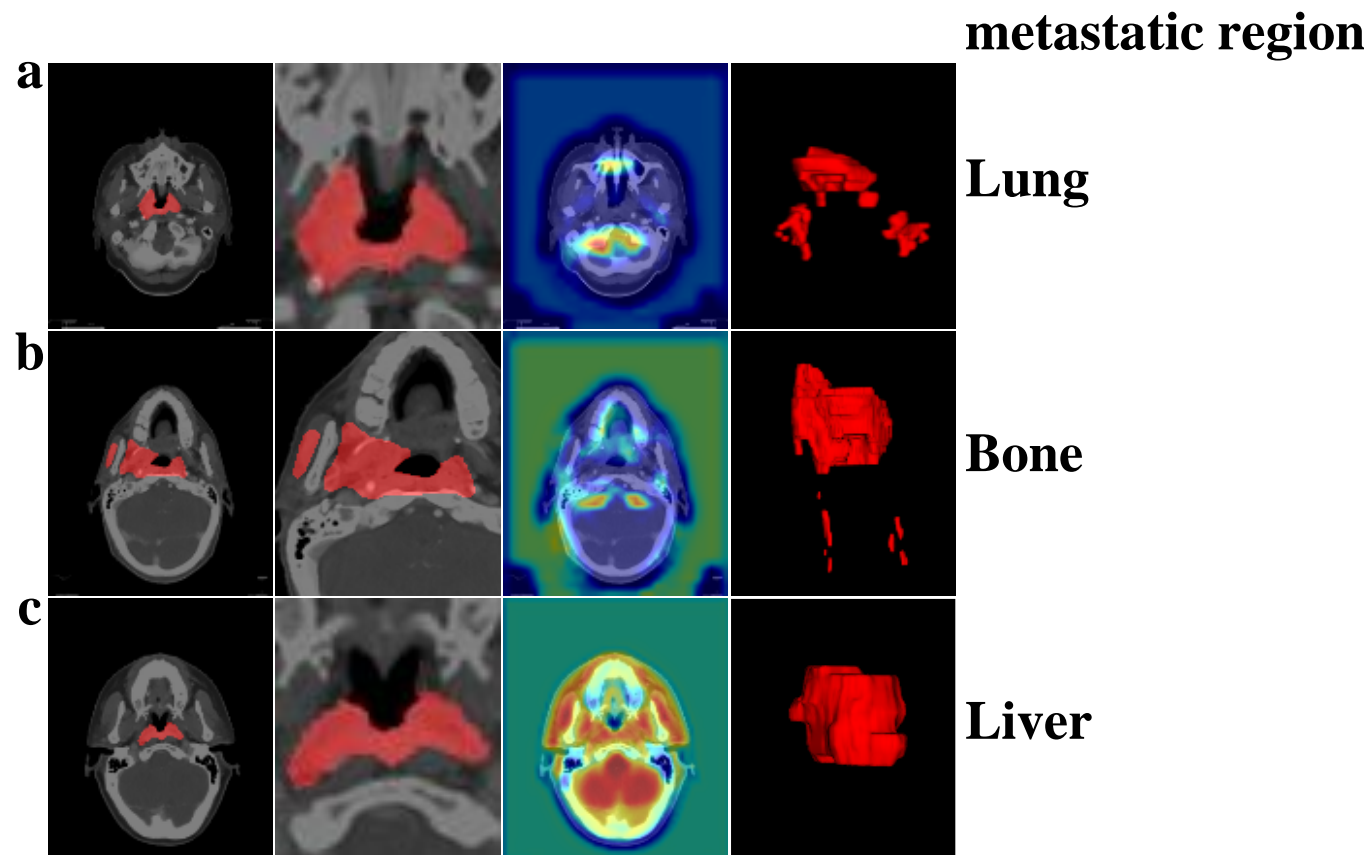

Figure 6: Representative heatmaps of incorrectly predicted HNSCC patients with distant metastasis were generated using Grad-CAM. The visual presentation was organized into four columns: displaying the CT images alongside the enlarged tumor regions, corresponding heatmaps, and 3D tumor volumes. The heatmaps were color-coded to highlight the discriminative regions and superimposed on the respective ultrasound images. It can be observed that the model attention did not accurately focus on the true tumor region, indicating potential areas for improvement.

# 6.[Reference]